\documentstyle[amsmath,amssymb,graphicx]{article}

\def\be{\begin{eqnarray}}
\def\ee{\end{eqnarray}}
\def\nn{\nonumber}

\def\p{\partial}

\def\Tr{{\rm Tr}\,}

\voffset= - 1.25in
\hoffset= - 1.0in         

\begin{document}


\hfill ITEP/TH-6/13

\bigskip

\centerline{\Large{Integrability in non-perturbative QFT\footnote{
Talk at the
2nd International Workshop on Nonlinear and Modern Mathematical Physics
March 9-11, 2013, Tampa, Florida
  }
}}

\bigskip

\centerline{A.Morozov}

\bigskip

\centerline{\it ITEP, Moscow, Russia}

\bigskip

\bigskip

\centerline{ABSTRACT}

\bigskip

{\footnotesize
Exact non-perturbative partition functions of coupling constants
and external fields exhibit huge hidden symmetry,
reflecting the possibility to change integration variables
in the functional integral.
In many cases this implies also some non-linear relations
between correlation functions, typical for the tau-functions
of integrable systems.
To a variety of old examples,
from matrix models to Seiberg-Witten theory
and AdS/CFT correspondence,
now adds the Chern-Simons theory of knot invariants.
Some knot polynomials are already shown to combine into tau-functions,
the search for entire set of relations is still in progress.
It is already known, that generic knot polynomials fit into the set of
Hurwitz partition functions -- and this provides one more stimulus
for studying this increasingly important class of deformations
of the ordinary KP/Toda $\tau$-functions.
}

\bigskip

\bigskip

\section{Introduction}

For a long time integrability was thought to be
an exotic phenomenon, useful for providing
solvable examples in the textbooks and
for enjoying the intellectual gourmands
by mathematical beauties
and possibilities to tie together the different branches of science.

However, today, despite the growing importance
of these motivations, especially the last one --
actually, the key goal of the {\it string theory} \cite{UFN2}, --
integrability occupies a much bigger place in
theoretical studies, and its role and significance
is only increasing.
The word "integrability" appeared hundreds of times
only in the titles of the hep-th papers during last several years,
well known subjects include matrix models,
Seiberg-Witten theory of non-perturbative Yang-Mills fields,
AdS/CFT-correspondence and many newly-emerging topics --
Hurwitz and knot theories among the most recent ones.
Ironically, when one performs this search in arXiv,
it lists together the titles with "integrability"
and "integrals", predominantly functional.
There it is, probably, just the linguistics,
but in fact this is symbolic,
because integration is indeed the place,
where the modern role of integrability has its origin.

In fact, today we think \cite{UFN2} that integrability is the pertinent
feature of non-perturbative (i.e. exact) partition functions --
the quantities obtained by evaluation of functional
integrals, and depending on the parameters of the integration
measure (coupling constants), boundary conditions
(background fields and vacuum condensates)
and the integration domain (cut-offs).
The main property of the integral is that it does not depend
on the integration variables -- and this is the property,
which is finally reflected in integrability features of
the partition function, often not immediately visible (hidden),
when it is calculated by various techniques, which (most of them)
do not explicitly respect integrability.

\section{Renormalization group}

Traditional way to think about integrable systems is in terms
of commuting Hamiltonian flows on the phase space --
but this is not at all straightforward to see and identify
these flows in the study of non-perturbative physics.
Perhaps, the most obvious place where at least {\it one} flow
is studied for many years in this context, is the
renormalization group.

Namely, when one tries to define a partition function like
\be
Z(g_k|\Lambda) = \int_{|p|<\Lambda} D\phi(x) e^{iS\{\phi\}/\hbar}
\ee
where, morally,
\be
S\{\phi\} = \int \Big((\partial\phi)^2 +  \sum_k g_k\phi^k\Big)d^dx
\ee
it is possible to ask that the coupling constants $g_k$
depend on the cut-off $\Lambda$ in such a way, that
\be
\frac{d}{d\Lambda}Z\Big(g_k(\Lambda)\Big|\Lambda\Big) = 0
\ee
This {\it renormalization procedure} is extremely interesting
by itself, and itself involves a lot of group theory and
integrable-like structures \cite{DK,GMS}
-- but this is beyond the scope of this text.
What matters, this principle defines a flow
\be
\frac{\partial g_k}{\partial\Lambda}=\beta_k(g)
\ee
on the space of the coupling constants.
Of course, it is just a single flow, to make it into
a set of flows one should consider arbitrary deformations
of the boundary (cut-off) of the integration domain \cite{Pol,MiMoRG},
what was never  really studied in quantum field theory.

Still, even looking at one flow, one can say something.
The point is that the space of coupling constants is
multi-dimensional, and in such case a generic flow is
"chaotic". This is what one {\it could} expect from generic
renormalization group \cite{MN}, but this is what was yet
never observed in any QFT example.
Partly this is reflected in various versions of
Zamolodchikov's $c$-theorem \cite{Cth},
but actually the statement is stronger:
the flow could be contractive, but still chaotic in orthogonal
directions.
The fact that this does not seem to happen,
can be a signal of a hidden integrability --
and in a context, much broader than any of the already
well-studied examples.

\section{Ward identities and AMM/EO topological recursion}

Still, the flows are not the most convenient language
to describe integrability in modern days.
It usually shows up differently:
as an infinite set of consistent differential equations
on partition function $Z(g_k)$.
These equations are in fact nothing but Picard-Fuchs equations for
the integrals, which depend on parameters -- and they reflect
the possibility to change integration variables,
what can be alternatively imitated as the change of
the coupling constants.
In QFT such equations are also known as Ward identities.

The archetypical example \cite{vircoMM} is the matrix integral
(matrix model) -- with already many enough integration variables
to observe non-trivial structures, and with still few enough
to make as detailed calculations as one can wish.
This calculation leads to the celebrated {\it Virasoro constraints}
{\it a la} \cite{virco}:
\be
\hat L_n Z(t) = \sum_k k(T_k+t_k)\frac{\p Z}{\p t_{k+n}}
+ \hbar^2\sum_{a+b=n}\frac{\p^2 Z}{\p t_a\p t_b}
= 0, \ \ \ \ n\geq -1
\label{virco}
\ee
which express invariance of
\be
Z(t_k) = \int_{N\times N} dM \
e^{ \frac{1}{\hbar}\left(-\Tr W(M) + \sum_{k} t_k\Tr M^k\right)}
\label{Mamoint}
\ee
under the change $\delta M = \epsilon M^{n+1}$.
This allows to define the partition function as a $D$-module,
for which (\ref{Mamoint}) is just an integral representation.
As to the integral, it can be well defined by introduction of background potential
$W(z) = -\sum_k T_kz^k$, usually polynomial, for example, quadratic.

Amusingly, the dependence on this, originally auxiliary,
potential appears to be one of the most interesting \cite{AMM}.
The terms of the {\it genus expansion} of the logarithm $\log Z$
of the partition function in powers of t'Hooft coupling constant $\hbar^2N$,
are expressed as the meromorphic poly-differentials
on a hyperelliptic Riemann surface,
\be
y^2 = W'(z)^2 - f(z)
\ee
defined by the polynomial $W(z)$ and some
additional data $f(z)$ -- hidden in the naive definition (\ref{Mamoint}).
Virasoro constraints then turn into a recurrent set of relations,
actually building a set of poly-differentials on the families of Riemann
surfaces (often called {\it spectral curves}),
which is now known under the name of
AMM/EO topological recursion \cite{AMM,AMM/EO} and which is now being
found in a variety of different fields, from Seiberg-Witten theory
of instanton sums \cite{KM,SW,LMNS} to knot theory \cite{FD}.
From integrability-theory point of view this phenomenon
-- emergence of a new recursions from expansions of a $D$-module --
is an example of emerging Whitham hierarchies,
to be briefly  described in s.\ref{Whith} below.

In $M$-theory context the spectral curve can be nicely visualized as
a surface, on which the $6d$ theory is compactified \cite{brane6d},
and this picture leads to various deep insights \cite{Gai},
including the AGT conjecture \cite{AGT}.

\section{Group theory and quadratic relations}

Ward identities are {\it linear} relations, imposed on exact
non-perturbative correlators.
Topological recursion looks quadratic, but this is only
a result of rewriting the action of quadratic operator
on the exponential -- the underlying equations are still
essentially linear (though this is not so easy to see in
some recent applications of the topological recursion).

Remarkably, non-perturbative partition functions
can also satisfy {\it truly} quadratic relations,
which have nothing to do with linearity.
They are usually called Hirota-like equations,
and functions, that satisfy such equations,
are known as $\tau$-functions. The best known is
KP $\tau$-function, satisfying
\be
\oint dz\ e^{\sum_k(t_k-t'_k)z^{-k}}
\tau\left(t_k +\frac{z^k}{k}\right)
\tau\left(t_k'-\frac{z^k}{k}\right)=0
\label{Hir}
\ee
what, if expanded in powers of $t_k-t'_k$, provides
an infinite set of quadratic differential equations
(KP hierarchy):
\be
3(\tau_{22}\tau - \tau_2^2)-4(\tau_{13}\tau-\tau_1\tau_3) +
(\tau_{1111}\tau-4\tau_{111}\tau_1 + 3\tau_{11}^2) =0, \nn \\
\ldots
\ee

The origin of such equations \cite{GKLMM} is in the theory of Lie groups,
where one can multiply representations
\be
R_1\otimes R_2 = \oplus_i R_i
\ee
i.e. there is a comultiplication and intertwining operators.
Quadratic relations appear when one consider two different
decompositions with some common representation $Q$,
belonging to both:
\be
R_1\otimes R_2 \longrightarrow Q \longleftarrow R_3\otimes R_4
\ee
Then
a system of quadratic relations
can be written
on matrix elements of the group elements \cite{MV},
in different representations --
and this will be a Hirota-like equation for a generating function
of such matrix elements.

Unfortunately, not too many examples are worked out so far,
beyond the KP/Toda family, which is associated to totally
antisymmetric representations of the $\widehat{U(1)}$ algebra.
Remarkably, a new interest is now emerging to a quasiclassical
limit of this construction for quantum groups -- it is related
to cluster algebras and to the Seiberg-Witten theory \cite{Mars}.

A very interesting fact about the tau-functions is that they
can often be described in terms of auxiliary Riemann surfaces
-- spectral curves. In other words, the Liouville tori of
integrable systems are often Jacobians of Riemann surfaces,
and complex moduli are invariants of the Hamiltonian flows.
It is still not clear, how general this phenomenon is,
but it definitely holds for KP/Toda families \cite{Krich}.

\section{Interplay between the linear and quadratic  relations}

Thus non-perturbative partition functions naturally satisfy
rich sets of {\it linear} Ward identities -- as a result of the freedom
to choose integration variables in the functional integral --
and {\it quadratic} relations -- as a result of existence of
operator formalism in quantum mechanics, where amplitudes
are matrix elements.
With this understanding of their origins,
it is clear, that the connection between linear and
quadratic relations is not going to be very straightforward --
as is the case of the subtle connection between the two formalisms
in quantum theory.
Indeed, even relation between the Virasoro constraints
(\ref{virco}) and KP Hirota equation  (\ref{Hir}) for the
simple matrix model (\ref{Mamoint}) is not fully understood.

\bigskip

In any case, it is clear that non-perturbative partition functions
should be some kind of the tau-functions, but definitely not
of generic type: in addition to {\it quadratic} they should satisfy
at least one {\it linear} constraint -- then all other
Ward identities will follow.
This single constraint is often called {\it string equation},
and such solutions to Hirota-like equations are called
{\it matrix-model $\tau$-functions} \cite{UFN3}.

\section{Whitham hierarchies and Seiberg-Witten theory
\label{Whith}}

Normal in quantum field theory is not to define partition functions
by integrating over all the fields completely, but rather do so
with short-distance fluctuations, so that the answer remains
dependent on the background slowly-varying fields, and effective
action contains only terms with few derivatives.
Such procedure gets especially important, because we believe
that the Standard Model of the fundamental interactions arises
in exactly this way from some still unknown UV-complete theory
at Planckian scale (the superstring model \cite{sust} being one
of the most popularized candidate, provided in the string theory
context \cite{UFN2}).
This procedure has, of course, a counterpart
in integrability theory, where averaging over fast variables
converts ordinary integrable hierarchies, satisfying Hirota
equations, into effective theories of slow modes,
described by the {\it Whitham theory} \cite{Whith}.
The crucial feature of Whitham hierarchy is that it depends
on the choice of solution to original (high energy) hierarchy,
and in the case when those solutions are parameterized by
spectral curves,-- it depends on the spectral curve.

In other words, the belief is that the low-energy effective
actions should be somehow described in terms of some
additional hidden structure -- auxiliary Riemann surfaces,
or, better, the families of such surfaces, parameterized
by the choice  of the vacuum.
Whitham dynamics defines the dependence of the low-energy
effective action on the moduli -- and the result is that
it is often constrained by a new kind of non-linear equations:
the {\it WDVV-like system} \cite{WDVV}
\be
{\cal F}_i{\cal F}_j^{-1}{\cal F}_k = {\cal F}_k{\cal F}_j^{-1}{\cal F}_i
\ee
where $\big({\cal F}_i\big)_{jk} = \p^3_{ijk} \log {\cal Z}$
are matrices, made out of the third derivatives of the
low-energy {\it prepotential}  ${\cal F} = \log {\cal Z}$, and derivatives
are with respect to some special coordinates $\{a_i\}$ on the moduli
space of spectral curves, introduced via the Seiberg-Witten
(or special-geometry) equations
\be
\oint_{A_i} \Omega = a_i, \ \ \ \ \ \ \ \
\oint_{B_i} \Omega = \p _i{\cal F} = \frac{\p\log {\cal Z}}{\p a_i}
\ee

This construction appeared to be extremely successful \cite{GKMMM}
in application
to the only known explicit example of exact non-perturbative
effective action: the one obtained by summation over instantons
in the $N=2$ supersymmetric Yang-Mills theory in 3,4,5 and 6
space time dimensions \cite{SW,LMNS}.
In fact, it establishes a one-to-one correspondence between various
quiver gauge theories and 1-dimensional integrable systems, like
Lioville/Toda systems and spin chains, provides an effective
description of $S$-dualities \cite{Gai,GMM} and relates them to
non-trivial spectral dualities between integrable systems, like
\cite{AHH,MMRZZ}.

\section{Lifting Whitham prepotentials to $\tau$-functions}

Of course, low-energy effective action is
only a small remnant of original non-perturbative partition function.
Likewise the Whitham prepotential is only a small remnant of original
$\tau$-function, and there can be many different non-perturbative
partition functions, leading to the same prepotential.
It can be natural to look for the {\it minimal} among such extensions.

In the case of the Seiberg-Witten theory, where Whitham
dynamics is associated with entire world of 1-dimensional
integrable systems, such an extension should be also
distinguished.
Today we already know that the corresponding minimal
partition functions are closely related with conformal blocks
of $2d$ conformal field theories -- this is the celebrated
AGT relation \cite{AGT,MMMM}.
The best possible formulation of it is in terms of
Hubbard-Stratanovich duality in a peculiar Dotsenko-Fateev
matrix model \cite{DFmamo},
where Seiberg-Witten spectral curve and associated
genus expansions are naturally appearing.
A particular subset (particular Nekrasov-Shatashvili limit)
of these conformal blocks is related to quantization
of the $1d$ integrable systems \cite{NSL}.
$S$-duality relations seem to be realized in terms of $3d$
Chern-Simons theory.

\section{Chern-Simons theory and knot polynomials}

After distinguished $1d$ and $2d$ quantum field theories
found their place in the world of universality classes
of non-perturbative partition $\tau$-functions,
the $3d$ Chern-Simons theory \cite{CST}
is the natural next model to consider.
And indeed the growing efforts are now applied in this direction.

Chern-Simons theory is a version of Yang-Mills theory,
with a topological action:
\be
\int D\!{\cal A}\ e^{\frac{1}{\hbar}\int_{{\cal M}_3}
({\cal A}d{\cal A} + \frac{2}{3}{\cal A}^3)}
\ee
and the observables are gauge invariant Wilson-loops
\be
\Tr_R\ P\!\exp \oint_{\cal K}{\cal A}
\ee
along various lines (knots) ${\cal K}$ embedded into
a three-dimensional manifold ${\cal M}_3$.
The Wilson-loop averages are often called {\it knot functions},
they depend on the space ${\cal M}_3$,
on the knot (or the link) ${\cal K}$,
on the gauge group $G$, on its representation $R$
and on the coupling constant $\hbar$.
For the simply connected ${\cal M}_3=R^3$ or $S^3$
and for $G=Sl(N)$ the knot functions are actually
{\it polynomials} of the variables $q = e^{2\pi i \hbar}$
and $A = q^N$ -- and they are often called
{\it HOMFLY polynomials}:
\be
H_{\otimes_i R_i}^{\cup_i{\cal K}_i\in{\cal M}_3}(q|A) =
\left< \prod_i \Tr_{R_i}\ P\!\exp \oint_{{\cal K}_i}{\cal A} \right>_{CS}
\ee

There are different ways to carefully define and evaluate
HOMFLY polynomials, leading to various kinds of important quantities
and structures, see \cite{TR} for classical results and
\cite{knots} for relatively fresh reviews.
In particular, if knots/links are represented by the closure of braids
in the projection to the two-dimensional plane
(this happens if the functional integral is evaluated in the temporal
gauge $A_0=0$),
then knot polynomials exhibit a character decomposition
\cite{RJ,BEM,DMMSS,MMMknII}:
\be
H_R^{\cal K}(q|A) = \sum_Q C_{RQ}^{\cal K}(q) \chi_Q(q|A)
\ee
which completely separates knot (${\cal K}$) and group ($A$) dependencies.
In the case of HOMFLY polynomials the expansion basis $\chi_Q$
is provided by Schur functions
(the characters of $Sl(\infty)$), and this implies a natural
deformation to MacDonald polynomials -- such deformation
of knot polynomials, depending on two parameters $q$ and $q'$,
indeed exists and leads \cite{DMMSS,Che} to {\it superpolynomials}
\cite{DGR,sups}.
Character expansion is actually defined on a bigger space
of time variables instead of just $A$, where one introduces
{\it extended knot polynomials}   \cite{DMMSS,MMMknI},
\be
{\cal P}_R^{\cal K}(q,q'|t_k) = \sum_Q C_{RQ}^{\cal K}(q,q') \chi_Q(q,q'|t_k)
\label{extpols}
\ee
which can be directly compared to $\tau$-functions.
It turns out that
-- after appropriate summation over representations $R$ --
they indeed give rise to $\tau$-functions
in the case of the {\it torus knots} \cite{MMMknI},
but for generic knots this is not true: some modification
is needed, either of the extension of knot polynomials,
or of the $\tau$-functions, perhaps, taking into account
the existence of "quantum" parameter $q$.


From the point of view of integrability, the most interesting
seems the {\it genus expansion} near the point $q=1$.
Exactly at this point the HOMFLY polynomial exhibits a very simple
dependence on the representation:
\be
H_R^{\cal K}(q=1|A) \ =\ \Big(\sigma_\Box^{\cal K}(A)\Big)^{|R|}
\ee
where $|R|$ is the number of boxes in the Young-diagram $R$,
and  $\Box$ denotes the fundamental representation, represented
by a diagram with a single box.
However, genus expansion is far more interesting \cite{MMS}:
the first correction is
\be
H_R^{\cal K}(q|A) \ =\ \Big(\sigma_\Box^{\cal K}(A)\Big)^{|R|}
+ (q-q^{-1})\varphi_R([2])\Big(\sigma_\Box^{\cal K}(A)\Big)^{|R|-2}
{\sigma_{[2]}^{\cal K}(A)} + \ldots
\ee
and in general
\be
H_R^{\cal K}(q|A) = \Big(\sigma^{\cal K}_\Box(A)\Big)^{|R|}
\exp \left(\sum_j (q-q^{-1})^{^j}\!\!\!\sum_{|Q|\leq j+1}\varphi_R(Q)\,
\frac{\sigma^{\cal K}_{j|Q}(A)}{\big(\sigma^{^{\cal K}}_\Box(A)\big)^{2j}}\right)
\label{geHOMFLY}
\ee
Here $\sigma_{j|Q}(A)$ are various {\it special polynomials}
in $A$, with coefficients made out of {\it Vassiliev invariants}
\cite{Vasinv} of the knot ${\cal K}$.
The $R$-dependent coefficients $\varphi_R(Q)$ are the
characters of symmetric group $S_\infty$ --
which are nowadays studied in the framework of the {\it Hurwitz theory}.
Note that for a given order $j$ of the genus expansion
only Young diagrams of the sizes $|\Delta|\leq j+1$ contribute,
if one did not exponentiate the expansion, analogous restriction
would be weaker: $|\Delta|\leq 2j$.

Extremely interesting seems the generalization of genus expansion
(\ref{geHOMFLY}) to superpolynomials, of which only the first-order term
is conjectured \cite{AntMor,AMMM21}:
\be
P_R^{\cal K}(q,q'|A) = \Big(P_\Box^{\cal K}(q,q'|A)\Big)^{|R|}
+ \Big( (q-q^{-1})\,\nu_{R'} -(q'-q'^{-1})\,\nu_R\Big)
{\sigma_{[2]}^{\cal K}(A)}\left(\sigma^{\cal K}_\Box(A)\right)^{|R|-2}
+ \ldots
\label{suspe1}
\ee
where $\varphi_R([2]) = \nu_{R'} -\nu_R$, $R'$ is transposed diagram $R$,
and $\nu_R = \sum_j (j-1)r_j$.
It should also include a non-trivial deformation of Hurwitz theory,
describing {\it refinement} of symmetric group characters $\varphi_R(Q)$,
which still remains to be found.

An intriguing question is the consistency
between (\ref{suspe1}) and the desired positivity of the colored superpolynomial:
beyond non-(anti)symmetric representations the relation can be more
involved \cite{AMMM21}, while (\ref{suspe1}) can instead describe
MacDonald-inspired not-always-positive colored superpolynomials of \cite{DMMSS,Che}.

\section{Hurwitz partition functions}

The central object in Hurwitz theory \cite{MMN} is the set of commuting operators
$\hat W_R$, parameterized by Young diagrams $R=\{r_1\geq r_2\geq\ldots \geq 0\}$ --
labeling the conjugation classes of permutations of $|R|$ elements
(see \cite{MMNnc} for a non-commutative extension, associated with open
rather than closed string theory).
These operators have $Sl(\infty)$ characters (Schur functions)
as their common eigenfunctions, while the corresponding eigenvalues
are the symmetric group characters $\varphi_R(Q)$:
\be
\hat W_Q \chi_R = \varphi_R(Q)\chi_R
\ee
The simplest is representation of the operators in Miwa coordinates
$t_k = \frac{1}{k}\Tr X^k$, then
\be
\hat W_R \sim \ :\prod_i \Tr \left(X\frac{\p}{\p \tilde X}\right)^{r_i}\!:
\ee
In ordinary time-variables the first non-trivial is the
{\it cut-and-join operator}
\be
\hat W_{[2]} = \sum_{a,b} \left(abt_at_b\frac{\p}{\partial t_{a+b}}
+ (a+b)t_{a+b}\frac{\p^2}{\p t_a\p t_b}\right)
\ee
which appears in various branches of science, including
the construction of Khovanov knot polynomials \cite{KH}.

Relation to Hurwitz numbers (counting ramified $n$-sheet genus-$g$ coverings of
Riemann surfaces) is through the Frobenius formula \cite{Fro}
\be
{\rm Cover}_{n,g}(Q_1,\ldots,Q_p)
= \sum_{|R|=n} d_R^{2-2g}\varphi_R(Q_1)\ldots\varphi_R(Q_p)
\ee
It implies consideration of the generating functions of the form
(at genus $g=0$)
\be
Z(\beta|t) = \exp \left(\sum_Q \beta_Q\hat W_Q\right) e^{t_1}
= \sum_R \exp\left(\sum_Q \beta_Q\varphi_R(Q)\right) \chi_R(t)
\label{Hupf}
\ee
which we call {\it Hurwitz partition functions}.

It is clear from (\ref{geHOMFLY}) that HOMFLY polynomials
are actually the Hurwitz partition functions, evaluated at some
knot-dependent point in the space of $\beta$-variables.
They can actually be rewritten in terms of the operators $\hat W$,
if one considers the Ooguri-Vafa generating function
\be
Z_{OV}^{\cal K}(t) = \sum_R H_R^{\cal K} \chi_R(t)
\ee
-- note that this introduces a complementary set of times to (\ref{extpols}).
Then (\ref{geHOMFLY}) implies that
\be
Z_{OV}^{\cal K}(t) =
\exp \left(\sum_j (q-q^{-1})^{^j}\!\!\!\sum_{|Q|\leq j+1}
\frac{\sigma^{\cal K}_{j|Q}(A)}{\big(\sigma^{^{\cal K}}_\Box(A)\big)^{2j}}\, \hat W_Q\right)
\exp\left(\sum_k \frac{(A^k-A^{-k})\cdot
\Big(\sigma_\Box^{\cal K}(A)\Big)^k}{q^k-q^{-k}}\,t_k\right)
\ee
It was conjectured \cite{OV} that the character expansion of $\log Z_{OV}$
possesses non-trivial {\it integrality}
properties (not to be mixed with {\it integrability}, at least immediately),
which can be understood with the help of the theory of $\hat W$ operators
\cite{LP}.

Since operators $\hat W_Q$ form a {\it commutative} algebra,
$Z(\beta|t)$ -- and thus $Z_{OV}^{\cal K}(t)$,
as its particular case --
is clearly related to {\it integrable} systems.
However, only for the unknot
$Z_{OV}$ turns out to be an ordinary KP $\tau$-function,
satisfying Hirota equation (\ref{Hir})
and possessing a free-fermion representation.
For a general argument {\it against} such naive integrability for arbitrary
{\it knots} (but not {\it links}) see footnote 3 in \cite{MMMknI}.
In fact, integrability properties of Hurwitz partition function
are not fully understood
and their connection to KP/Toda integrability remains obscure.
In particular,
KP integrability in $t$-variables
is preserved only by the action of Casimir operators,
i.e. when the $\hat W_Q$ operators in (\ref{Hupf}) are {\it linearly} combined
to form integrability-preserving {\it Casimir operators},
what does not happen for arbitrary values of $\beta_Q$ \cite{inteCas}.
This can mean that either some change of $t$-variables is required
to make integrability visible, or that KP-integrability is
somehow deformed or generalized when one moves from $2d$ conformal
theories to $3d$ Chern-Simons.

\section{Conclusion}

Integrability proved to be an extremely important part of
quantum field theory: its main role is to describe
full non-perturbative answers in quantum field theory --
and this fact is deeply related to the very essence of
quantum theory, whether it is formulated in operator
or in functional integral formalism.
It, however, remains an open question, what is exactly
the right version of integrability theory which matters.
Absolute majority of the known examples involves the
simplest KP/Toda integrability, associated with the free fermion
(antisymmetric) representations of the group $\widehat U(1)$,
though traces of more general quantum groups are already seen.
Today the frontline seems to be at the study of Hurwitz
partition functions, of which knot polynomials provide
a vast set of examples: it is crucially important to
understand and effectively describe their properties.
Despite not immediately belonging to the set of KP/Toda
$\tau$-functions, they seem to be at least their closest
relatives, moreover, they seem to possess genus expansions,
described in terms of the spectral curves and
AMM/EO topological recursion.

\section*{Acknowledgements}

I am indebted to my numerous coauthors, especially to A.Mironov,
for thinking together about the issues, mentioned in this text.
My work is partly supported by
the Russian
Ministry of Education and Science
under contract 8498,
by NSh-3349.2012.2,
by RFBR grants 13-02-00478
and by the joint grants
13-02-90459-Ukr-f-a,
12-02-92108-Yaf-a,
13-02-91371-St-a
and 13-02-90618-Arm-a.
Special thanks are to Wen-Xiu Ma and Razvan Teodorescu
and to other organizers of the
Tampa Workshop
for the invitation and support.

\end{document}